\documentclass[12pt, reqno]{amsart}
\usepackage {amsfonts}
\usepackage{amsmath} 
\usepackage{amssymb} 
\usepackage{amsthm}
\usepackage{graphicx} 
\usepackage{epstopdf}
\usepackage{mathrsfs}
\usepackage{ifthen}
\usepackage{color}
\usepackage{tikz}
\usetikzlibrary{matrix,arrows}
\usepackage{xr-hyper}
\usepackage{wrapfig}
\usepackage{titlesec}
\usepackage{multicol}
\usepackage{url}
\usepackage{color}
\usepackage{hyperref}
\definecolor{refcolor}{RGB}{0,0,190}
\hypersetup{
    colorlinks,
    citecolor=refcolor,
    filecolor=refcolor,
    linkcolor=refcolor,
    urlcolor=refcolor
}

\definecolor{scolor}{rgb}{0.0,0.0,0.7}
\definecolor{bcolor}{rgb}{0.0,0.0,0.7}
\titleformat{\section}
{\color{scolor}\normalfont\Large\bfseries}
{\color{scolor}\thesection}{1em}{}
\titleformat{\subsection}
{\color{scolor}\normalfont\large\bfseries}
{\color{scolor}\thesubsection}{1em}{}

\definecolor{DarkRed}{rgb}{0.8,0.0,0.0}

\definecolor{quotecolor}{rgb}{0.15,0.0,0.66}
\newcounter{questioncounter}

\definecolor{mytitlecolor}{rgb}{0.20,0.0,0.95}

\newcommand{\image}[3]{\begin{figure*}[ht]
\includegraphics[width=#2\textwidth]{img/#1}
\caption{\small{\label{#1}#3}}\end{figure*}}

\newcommand{\imagenocap}[2]{\begin{figure*}[ht]
\includegraphics[width=#2\textwidth]{img/#1}
\end{figure*}}

\theoremstyle{definition}

\def\({\left(}
\def\){\right)}

\newcommand{\citep}[2]{\cite{#1}, p. #2}

\newcommand{\flrw}{Friedmann-Lema\^itre-Robertson-Walker}
\newcommand{\FLRW}{FLRW}
\newcommand{\schw}{Schwarzschild}

\newcommand{\schrod}{Schr\"odinger}

\begin{document}



\begin{center}
\href{http://fqxi.org/community/essay/winners/2013.1}{Fourth prize} in the FQXi's 2013 Essay Contest \href{http://fqxi.org/community/forum/category/31419}{\textbf{`It from Bit, or Bit from It?'}}
\\
This version appeared as a chapter in the collective volume {\em It From Bit or Bit From It? Springer International Publishing, 2015. pages 51-64}.
\end{center}

\begin{center}
\huge{\textbf{\color{scolor}\textit{The Tao of It and Bit}}
\\ \small{\em To J. A. Wheeler, at 5 years after his death.}}
\end{center}
\begin{center}
\textit{Cristi \ Stoica}
\\\footnotesize{\textit{E\,m\,a\,i\,l\,:\, \href{mailto:holotronix@gmail.com}{h\,o\,l\,o\,t\,r\,o\,n\,i\,x\,@\,g\,m\,a\,i\,l\,.\,c\,o\,m}}}
\end{center}


\begin{quote}
\textbf{Abstract.}
The main mystery of quantum mechanics is contained in Wheeler's delayed choice experiment, which shows that the past is determined by our choice of what quantum property to observe. This gives the observer a {\em participatory role} in deciding the past history of the universe. Wheeler extended this participatory role to the emergence of the physical laws ({\em law without law}). Since what we know about the universe comes in yes/no answers to our interrogations, this led him to the idea of {\em it from bit} (which includes the participatory role of the observer as a key component).

The yes/no answers to our observations ({\em bit}) should always be compatible with the existence of at least a possible reality -- a global solution ({\em it}) of the {\schrod} equation. I argue that there is in fact an interplay between {\em it} and {\em bit}. The requirement of {\em global consistency} leads to apparently acausal and nonlocal behavior, explaining the weirdness of quantum phenomena.

As an interpretation of Wheeler's {\em it from bit} and {\em law without law}, I discuss the possibility that the universe is mathematical, and that there is a ``mother of all possible worlds'' -- named the {\em Axiom Zero}.
\end{quote}

\section{Wheeler}
\label{sec:Wheeler}

John Archibald Wheeler was, arguably, the most influential physicist since Einstein, contributing to radical insights in general relativity, quantum mechanics, quantum field theory, quantum gravity, to mention just a few domains. Much of this influence was done through his many brilliant PhD students.
~\footnote{\label{note_wheelers_students}From Wheeler's students, I will mention only a few who changed the face of physics: Richard Feynman, Hugh Everett III, Jacob Bekenstein, Warner Miller, Robert Geroch, Charles Misner, Kip Thorne, Arthur Wightman, Bill Unruh, Robert Wald, Demetrios Christodoulou, Ignazio Ciufolini, Kenneth Ford, and others, to whom I apologize for not mentioning.}.
Although I've never met him, I see him as a person who is willing to risk his reputation by allowing him and his students to develop ideas which apparently contradicted the very foundations of physics, as accepted in his time. He worked on radical (at least for that time) subjects like {\em wormholes}, {\em black holes}, {\em geons} (objects made just of spacetime, including a way to obtain the mass and the electromagnetic field as effects of the topology of spacetime \cite{MW57}), {\em wavefunction of the universe}, with the accompanying {\em end of time}, strange superpositions of different topologies in a {\em quantum foam}, {\em delayed choice experiments} which seem to imply that the observer affects the past \cite{Whe78,delayed2007,ionicioiu2011quantum_delayed_choice}.
Moreover, the initial conditions of the observed system have to depend on those of the measurement device \cite{Sto12QMb}.

He encouraged his students to challenge well established paradigms, with ideas like:
\begin{itemize}
	\item 
A particle goes from one point to another by following all possible paths, even if it goes faster than light, or even back in time \cite{Fey85}.
	\item 
Many features of Quantum Mechanics can be better understood if we admit that there are many worlds \cite{Eve57,Eve73}.
	\item 
Black holes have their own thermodynamics, including entropy \cite{bekenstein1972black}. When combining the effect discovered by another student of Wheeler, Bill Unruh, with the principle of equivalence, we obtain the Hawking(-Zel'dovich-Starobinski) radiation.
\end{itemize}

Imagine how a PhD student coming with one of the above-mentioned ideas would be perceived. Such theories, even nowadays, appear to many as taken from science fiction, if not from new-age pseudoscience. How such ideas, instead of being ridiculed, were even accepted as top science? I would thank Wheeler's courage for the new generation of Einsteins who appeared and changed the face of modern physics -- if one genuinely wants to find or foster new Einsteins \cite{smolin2005noneweinstein}, one has so much to learn from him. And when his former students became widely acknowledged, he modestly remained in the shadow.

These beautiful theories were well-developed, to derive qualitative and quantitative predictions. Many of them were experimentally confirmed, while others are still waiting, and some just stand as beautiful concepts, whose role is to explain phenomena, rather than predicting new ones. Some of his ideas are so visionary, that probably we will never be able to verify them completely by experiments. His proposal {\em it form bit} \cite{Whe78,wheeler1983recognizing,wheeler1988world,wheeler1989information,Whe98geons} combines in an amazing way his previous results, and those of his students. This makes the subject of this essay.

\section{It-sy Bit-sy Spider}
\label{sec:itsy_bitsy_spider}

Imagine a world in which there are three kinds of beings: spiders, flies, and dragonflies. Spiders eat flies and dragonflies, but unfortunately they can't fly, so to catch their food, they have to build webs. Imagine there are two kinds of webs, one kind can catch only flies, and the other one can catch only dragonflies. So far nothing weird.

Now imagine spiders can see the prey flying, but their sight is not as good to detect what kind the prey is. They only see that whenever an insect flies towards a web, it is caught.

Spiders are very intrigued, because they wonder:

\begin{quote}
What we catch in a web-for-flies, is always a fly. What if we replace in the last moment the web-for-flies with a web-for-dragonflies? Obviously, in this case we would catch a dragonfly. But how can the kind of the prey be decided by our choice of the web? Was the prey a fly, or a dragonfly, before being caught in the web?
\end{quote}

\image{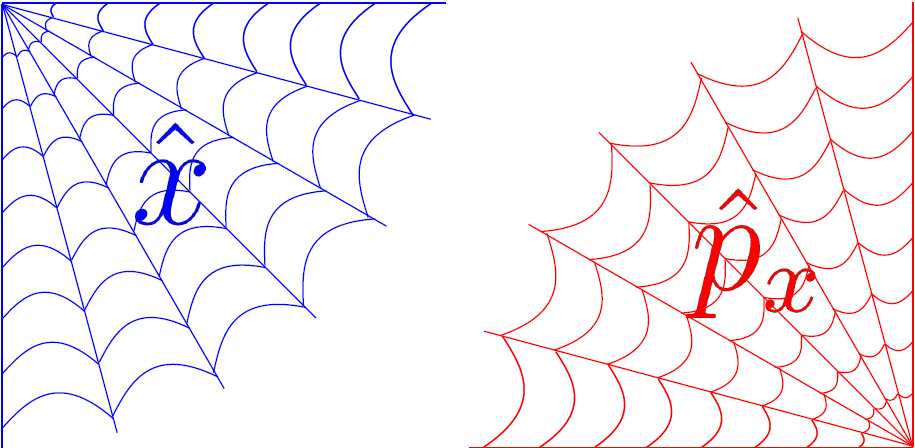}{0.7}{Spiderweb for catching particles, and spiderweb for catching waves.}

A quantum world is similar to a world in which the spider's choice of the type of the web determines what species is the insect which already flies toward the web.

Wheeler's {\em delayed choice experiment} can be seen as switching in the last moment the web with another kind of web, while the insect is still heading toward the web.

\section{Delayed Choice Experiment}
\label{sec:delayed_choice}

Recall the quantum experiment based on the Mach-Zehnder interferometer. Light is emitted by a source, and split by beam splitter $1$ (see fig. \ref{mz_both_ways}). The two halves of the ray are redirected by two mirrors to meet again, and the original ray is recomposed, by beam splitter $2$. The photons always trigger detector $B$.

\image{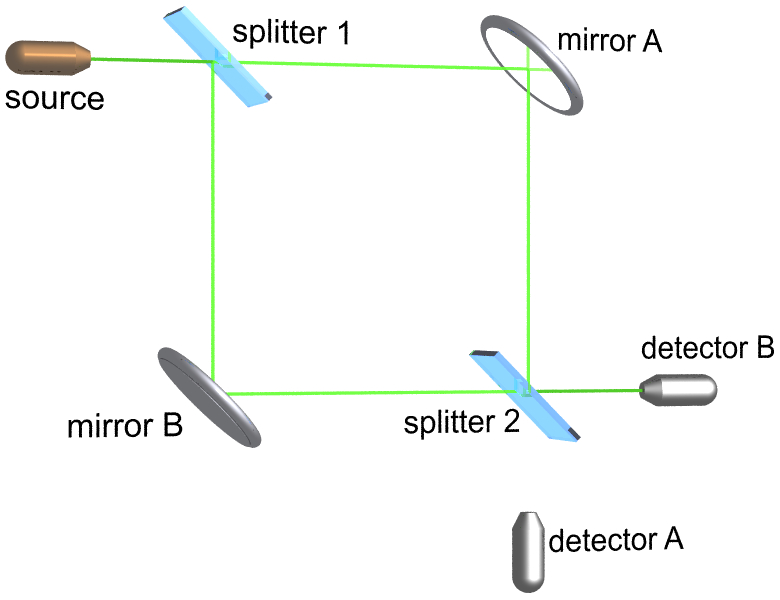}{0.49}{{\em Both ways} observation.}

Now, remove the beam splitter $2$. The photons will trigger with equal probability both detectors $A$ and $B$ (fig. \ref{mz_which_way}).

\image{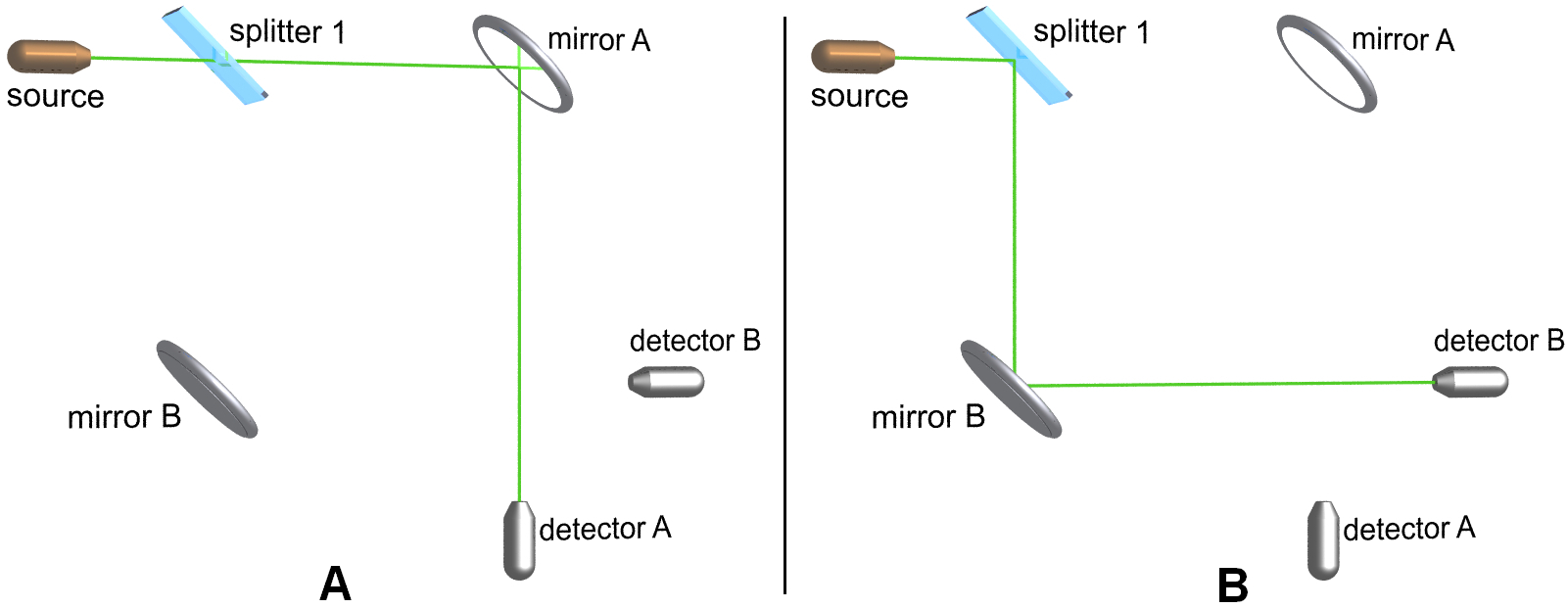}{0.98}{{\em Which-way} observation.}

Wheeler proposes to delay the decision of whether to keep or to remove the beam splitter $2$, until we are sure the photon passed from splitter $1$ \cite{Whe78}. In fact, his thought experiment uses instead of beam splitters and mirrors, the deflection of light caused by the gravity of an entire galaxy. He concludes \cite{Whe98geons}:

\begin{quote}
Since we make our decision
whether to measure the interference from the two paths
or to determine which path was followed a billion or so
years after the photon started its journey, we must
conclude that our very act of measurement not only
revealed the nature of the photon's history on its way to
us, but in some sense {\em determined} that history. The past
history of the universe has no more validity than is
assigned by the measurements we make--now!
\end{quote}

The delayed choice experiment is the source for Wheeler's {\em law without law} and {\em it from bit}.

\section{Law without Law}
\label{sec:law_without_law}

Wheeler pushed to the extreme his idea of delayed choice experiment. He thought that the observer determines not only the past of a quantum system, but the very physical laws! We can say that he extended his condensed formulation of Bohr's vision on quantum mechanics, ``no phenomenon is a phenomenon, until it is an observed phenomenon'', to ``no fundamental law is a fundamental law, until it is an observed fundamental law''.

Wheeler thought that the observer participates in choosing now the physical laws for the entire past and future history. He coined this vision {\em law without law}. He wrote in \cite{WZ83}

\begin{quote}
If the views that we are exploring here are correct, one principle, observer-participancy, suffices to build everything. [...] [The picture of the participatory universe] has no other than a higgledy-piggledy way to build law: out of the statistics of billions upon billions of acts of observer-participancy each of which by itself partakes of utter randomness.
\end{quote}

If Wheeler was right that we decide the physical laws, by our very choices as observers of the universe, then, due to their important and bold contributions to physics, he and his students are responsible for many preposterous features of our universe.

\section{Evolving Laws}
\label{sec:evolving_laws}

Regarding {\em law without law}, one may wonder how could there be different sets of laws to choose from. One possibility is that some constants are not really constants. They may became constant moments after the big-bang, frozen by symmetry breaking. Initially Wheeler proposed that after the big-crunch there will be a new universe, with different constants, but now we know that there will be no big-crunch \cite{PER99}.
A more recent proposal was made by Smolin, that the laws evolve from universe to {\em baby-universe} \cite{smolin1992did,smolin1997life,smolin2006status}. Presumably, a baby universe appears by going beyond a future spacelike singularity (like that of {\schw}). Penrose claims that this can't be done, because we can't match together a black hole and big-bang singularity, since they are of different types. They appear to be different, but there is an appropriate (singular) coordinate transformation which makes the {\schw} coordinate of the same type as the {\flrw} ({\FLRW}) one \cite{Sto11e}. In fact, at least in the case of the Oppenheimer-Snyder model of a black hole, the star is modeled as a time-reversed pure dust {\FLRW} solution, so it is not justified to claim that the two can't be matched together (a {\FLRW} singularity is the continuation of a time-reversed {\FLRW} singularity \cite{Sto11h,Sto12a}).

But Wheeler's philosophy {\em law without law} goes far beyond the idea of a mechanism of random mutations of the constants. He viewed the law as being created, or perhaps chosen from an infinity of alternatives, by the very observation process. The {\em bit} not only determines the (past) {\em it} of the universe, but also the laws.

\section{Tegmark's Mathematical Universe}
\label{sec:muh}

Tegmark's {\em mathematical universe} \cite{tegmark2008mathematical,Tegmark2014OurMathematicalUniverse} can provide an implementation of Wheeler's {\em law without law}. Tegmark proposes that all possible mathematical structures exist, and our universe is one of them.

He said that, in order for a universe to exist, it is enough to have a simulation of it, and that it is not even needed to run the simulation, merely having the description as a string of bits written on a CD-ROM is enough.

But the meaning of a string of bits depends on the language used to encode the information in it. The first comment I want to make about this is that the meaning of the string specifying our universe can be anything, including the specifications of any other possible universe, because for any possible meaning, one can always imagine a language in which the string has that meaning. Hence, any string, for example ``0'', is enough to specify all possible universes, given the appropriate decoding language.

The second comment is that the language has to be specified as well, in another language, and we arrive at an infinite regress. To avoid the regress, one can admit that there is a reality given by those specifications. Perhaps also an observer is needed,  a ``ghost in the quantum Turing machine'' \cite{Aaronson2013ghostQM}, something that ``breathes fire into the equations''\cite{Hawking1998ABriefHistoryOfTime}.


\section{It from Bit}
\label{sec:it_from_bit}

Wheeler tried to remove completely the idea of an independent reality ({\em it}), proposing that it emerges from the information contained in our observations ({\em bit}), which is the only one existent \cite{Whe98geons}:

\begin{quote}
it is not unreasonable
to imagine that information sits at the core of physics, just as it sits at the core of a computer
\end{quote}
and
\begin{quote}
I build
only a little on the structure of Bohr's thinking when I
suggest that we may never understand this strange
thing, the quantum, until we understand how
information may underlie reality. Information may not
be just what we {\em learn} about the world. It may be what
{\em makes} the world.
\end{quote}

Wheeler's \textit{it from bit} claims that the information is fundamental, more fundamental than anything else. But it is not simply a {\em digital theory of everything}. The central point is indeed the bit, the information about the universe, which is accessible to the observer. But equally important is the fact that the observer has a participatory role.

Wheeler often represented the universe as the letter $\mathbb U$, with the big-bang at the right end of the curve which makes the letter, and the observer at the left end, represented as an eye which, by mere observation, brings into existence the entire past history of the universe.

This is why Wheeler's {\em it from bit} should not be used to support the version of digital physics which just claims that ``everything is information''; nor should it be rejected by reducing his ideas to that idea \cite{barbour11bitfromit}. Wheeler made a much more profound point than that, as we have seen.

On the other hand, most of his arguments are based on the fact that we can only know bits of information, and on the delayed choice experiment. Besides the participatory role of the observer, which is difficult to deny, one should admit that the bits are subjective, pertaining to the observer. The fact that we can only collect bits of information doesn't really mean that there is nothing else but information.

\section{Can the Clicks of the Detectors Provide a Complete Description of Nature?}
\label{sec:clicks}

Is it possible to obtain {\em it} just from {\em bit}?


It is true that all quantum phenomena, no matter how weird they appear, are predicted by the very postulates of quantum mechanics. Strange behaviors such as correlations between the outcomes of measurements separated in space, and the fact that they depend on the context of the measurement, all follow from the simple postulates of quantum mechanics. Many try to find a more intuitive explanation for these phenomena, but they are simply explained by the fact that one can't simultaneously observe all properties of particles, because these properties are not well defined simultaneously \cite{Sto14quantumness}.

While it is undeniable that quantum mechanics is so successful, can we know everything about the universe just by quantum measurements? Can we even guess the physical laws from the outcomes of these measurements?

There is a big obstacle which prevents us for doing this. According to the postulates of quantum mechanics, the state of a system is represented by a vector in a complex vector space (the {\em state space} or the {\em Hilbert space}). But in a vector space there is no preferred basis, and the postulates of quantum mechanics are independent of any such basis. In reality, we know that the vector space containing all possible states has a richer structure, that the position in the physical space provides a preferred basis. We also know that each type of particle comes with its own state space, and the total space is obtained by taking tensor products between copies of these one-particle spaces. But these can't result simply by looking at the outcomes of quantum measurements, because the same outcomes would be obtained if the state vector of the universe is rotated in the state space by a unitary transformation. This shows that the information about the position basis and the tensor product structure is not encoded in the outcomes of measurements, so {\em it} doesn't simply follow from {\em bit} \cite{Sto14clicks}.

\section{Delayed Initial Conditions}
\label{sec:delayed_initial_conditions}

As a metaphor for the participatory universe, Wheeler mentions the game of twenty questions -- the player has to determine a word, by asking yes/no questions. The twist is that the word is not chosen at the beginning, but as the player asks the questions.

To make this work, the respondents have to take care that their combined answers still define a real word. This can only be done if they maintain, explicitly or not, a list of possible words. But it has to be at least one word on the list, at any moment.


What does this tell us about the universe? Classically, the state of the universe at any moment of time is determined by the initial conditions. This is prohibited in quantum mechanics, because we can only ask whether the system is in a small subset of possible states -- those particular states for which the property we measure is well-defined (fig. \ref{qm_observation}). It is not possible, even in principle, to know the complete state. There are no universal spiderwebs: each spiderweb can catch either flies, or dragonflies.

\image{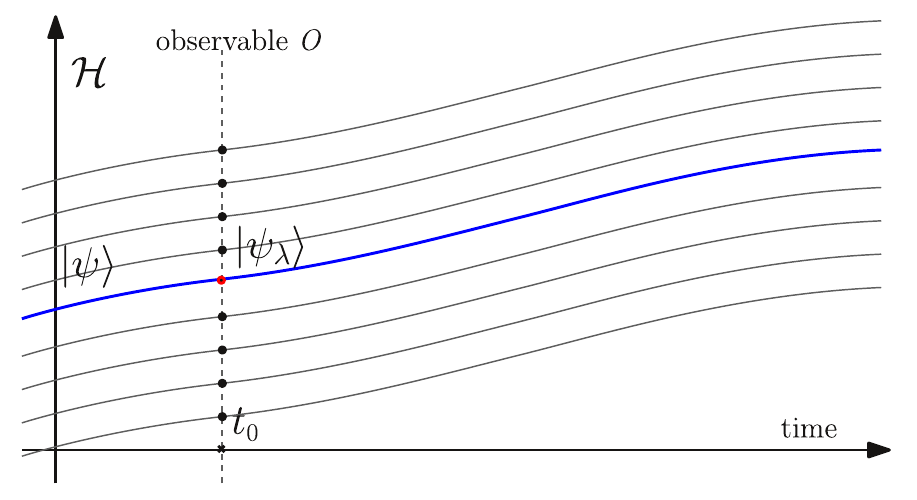}{0.7}{Any property we choose to observe, it is well-defined only for a small subset of the possible states. The observed system turns out to be in such a state.}

The observer asks questions, and the universe gives yes/no answers -- bits. But the answers always define at least a possible solution\footnote{If the initial conditions are fully specified, the solution is unique.
But our observations allow us to specify only partially the initial conditions, and that's why there are more possible solutions.}. It is not like there is no solution at all, as the catch-phrase {\em it from bit} implies.
Hence, one cannot infer that nothing exists, except the outcomes of the measurements. Rather, that at any given moment of time, there are possible realities which are compatible with those answers.

This is why I think that the complete picture is not {\em it from bit}, but rather {\em it from bit} $\&$ {\em bit from it}. The yes/no questions select a subset among the possible solutions of the {\schrod} equation, but the possible answers to the yes/no questions are determined by the possible solutions which remained (fig. \ref{determinism-delayed}) \cite{Sto08b,Sto08f,Sto12QMa}.

\image{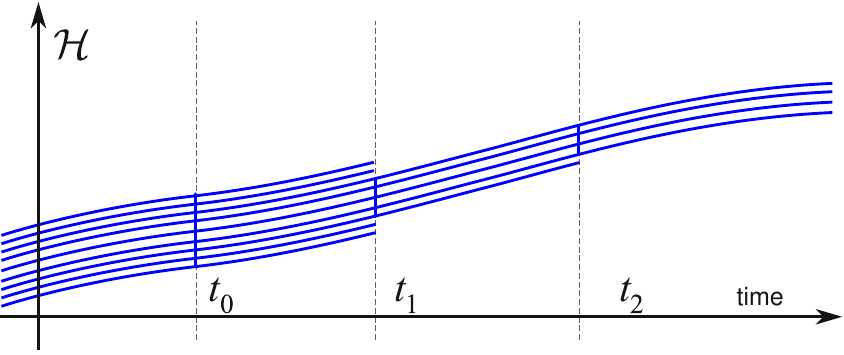}{0.7}{Delayed initial conditions select possible realities, even in the past.}

In addition, delayed initial conditions provide a way that free-will is compatible with deterministic laws \cite{Sto08e,Sto08f,Sto12QMa,Aaronson2013ghostQM}.

\section{Global Consistency Principle}
\label{sec:global_consistency}

Just because we don't have access to reality, but only to the bits, it doesn't mean that there is no reality. Which possibility is simpler: (1) that the yes/no bits are consistent with one another, that the probabilities are correlated, and that's all, or (2) that at any moment there is at least one possible reality, which ensure the consistency and the correlations? Isn't simpler and more logical the idea that {\em it} is something that prevents {\em bits} from contradicting one another, a ``reality check''.

Think at the way {\schrod} derived the energy levels from his equation. He had the equation, but he obtained the energy levels only after throwing away the solutions with bad behavior at infinity. The remaining solutions have, for an electron in an atom, a discrete spectrum. This provided the correct account to de Broglie's insight, that the wavelength of the electron's wave fits an integral number of times in the orbit. A global condition -- the boundary condition at infinity -- led him to the selection of only a discrete set of solutions from the continuum set of possible solutions of {\schrod}'s equation.

\image{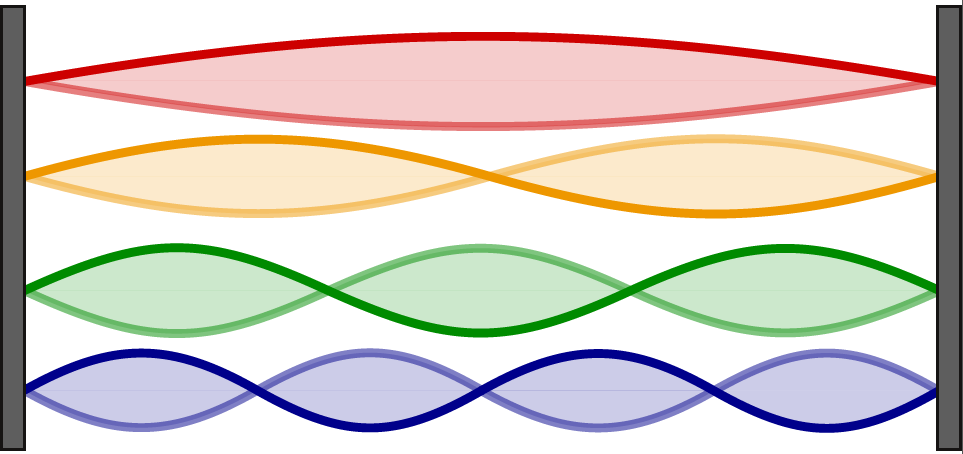}{0.5}{The role of global conditions.}

But how can the solution near an atom know how to be, so that it behaves well at infinity? This is a key question. If we think in terms of disparate bits, this can't hold in a natural way. If we think that the physical solutions have reality, it becomes natural to admit that they have to behave well at infinity (otherwise they can't have physical reality).

The {\em global consistency principle} generalizes the boundary conditions idea, and requires that no matter how are the observations spread in spacetime, there has to be a real solution for which the observations give the observed outcomes. For example, it requires that the presence or absence of the beam splitter $2$ in the experiment with the Mach-Zehnder interferometer has to be correlated with what happened with the ray at the beam splitter $1$.

\image{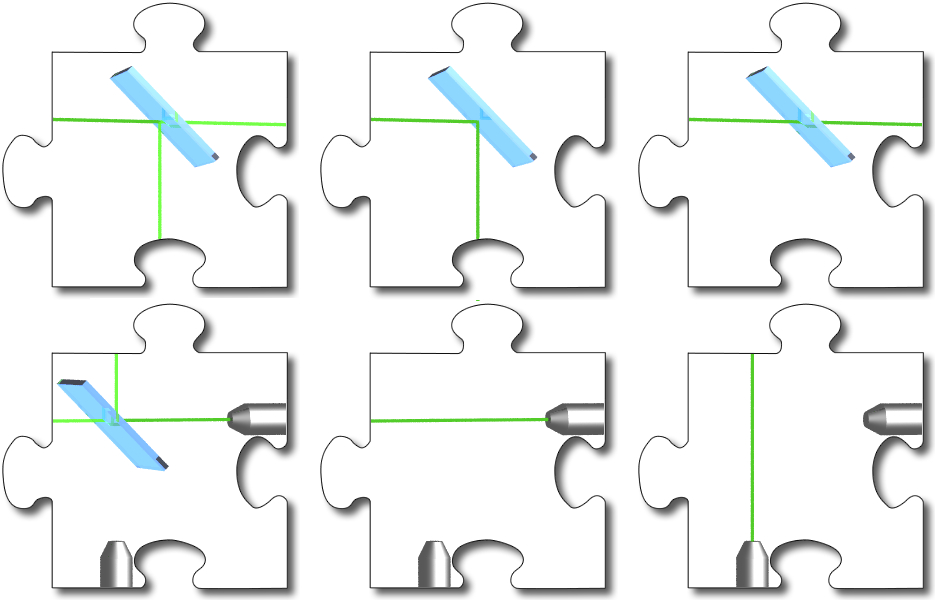}{0.7}{{\em Global consistency principle} requires that it has to exist a solution ({\em it}) which combines consistently all the pieces of the puzzle (the yes/no {\em bits} at different points and moments of time).}

To understand global consistency, it may help to remember that the solutions are defined on a four-dimensional spacetime, and to think in terms of an out-of-time view, like the block universe.

\section{The Big Book of the Universe}
\label{sec:big_book}

Here is why I find compelling the idea that our universe is mathematical. First, what we learn about anything, are {\em relations}. We don't know what water is, but we know its relation to our senses. Even its physical and chemical properties, follow in fact from interactions, hence from relations. Everything we know is defined by its relation with something else. If there is anything that can be mathematized, this is the {\em relation}. In fact, any mathematical structure is a set, along with a collection of relations defined between that set and itself \cite{birkhoff1946universalalgebra}.

\image{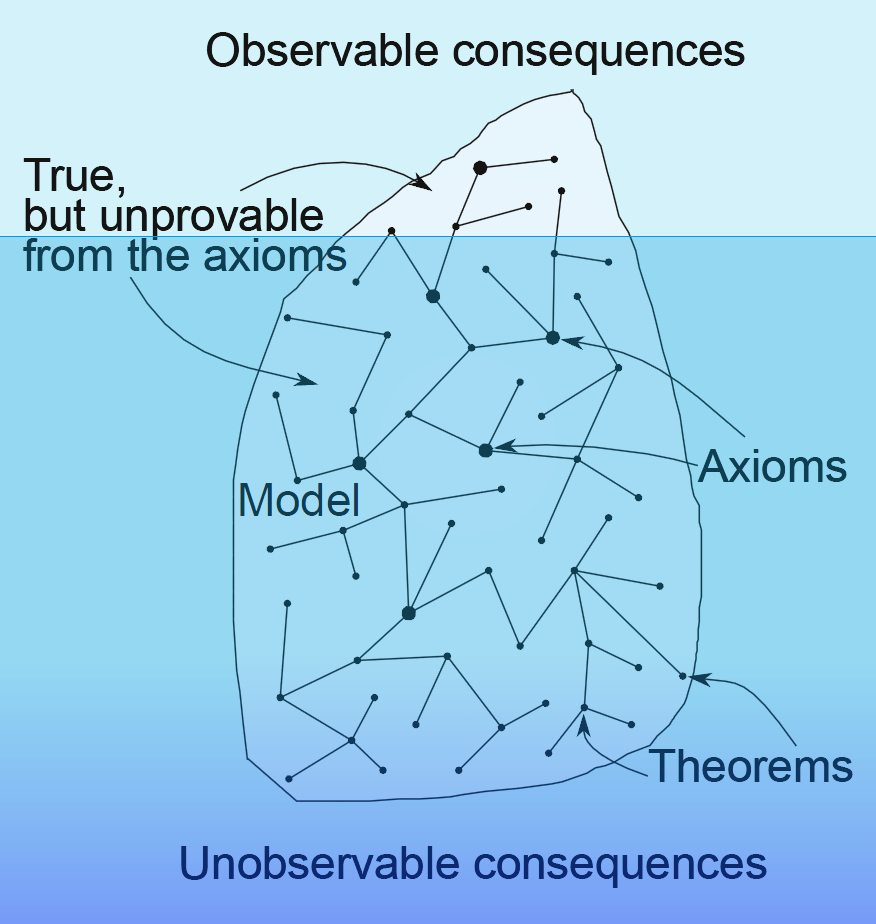}{0.6}{The iceberg represents the \textit{mathematical model} of the physical world. Its points represent true propositions, most of them unprovable from the axioms. The large dots represent \textit{axioms}, and the small dots consequences derived from them, or \textit{theorems}. The tip of the iceberg is what we can test by experiments and observations, at least in principle -- these are the \textit{observable consequences}. The largest part of the iceberg consists in untestable, or \textit{unobservable consequences}.}

Second, let's say that there is {\em a book containing every truth}. It will therefore contain the physical laws, and any truth about the state of a system at a given time -- the full description of the universe. Possibly the book is infinite. Maybe there is a finite subset of propositions in the book, from which everything else follows, or maybe not. G\"odel's theorem seems to say that there is no such finite subset, but maybe there is a finite subset from which everything follows by proofs of infinite length. Anyway, it seems very plausible that there may be a (possibly infinite) collection of propositions which contains all the truths about the universe. In this case, we have a a theory (of everything). To the theory we can associate a model, in the sense of {\em model theory} \cite{chang1990model}. A mathematical model is just a set with a collection of relations between its elements, a mathematical structure. So, whatever the collection of the truths about the universe is, the same propositions hold for that mathematical structure. The universe is isomorphic to a mathematical structure \cite{Sto13Math}.

``Wait!'', one may say, ``how about love, music, God, and so on? Are you claiming that these are just parts of a mathematical structure?'' Well, so long as these concepts are confined to a set of propositions, they are isomorphic to a mathematical structure. But what is wrong with this? For many, mathematics IS love, music, God... Maybe they have the ``fine ear'' for mathematics, maybe they hear in it the ``music of the spheres'' more than others, just like some have the ``fine ear'' for music.

Anyway, if one believes there are things that are not included in the mathematical model of the universe, one should describe those things. And this means that one has to build propositions about them, and to describe their relations with other things. And this means that they are already present in the book of all true propositions, and implicitly in the mathematical model.

\section{From Chaos to Law}
\label{sec:from_chaos_to_law}

One can go one step beyond {\em law without law}, and consider the following ``mother of all possible worlds''. Imagine a single axiom:

{\bf Axiom Zero.} {\em Axiom Zero} is false.

It is easy to see that from {\em Axiom Zero}, any possible proposition follows. Let's denote {\em Axiom Zero} by $p$. From {\em Axiom Zero} follows that its negation, $\neg p$, is also true. But from $p$ and $\neg p$, any proposition $q$ follows. This is known as {\em the principle of explosion}, or {\em ex falso quodlibet}. The proof that from contradiction anything follows is very simple~\footnote{Assuming both propositions $p$ and $\neg p$ are true, we want to prove $q$. Since $p$ is true, $p\vee q$ is true. But since $\neg p$ is true, $p$ is false. From $p\vee q$ and $\neg p$ follows that $q$ is true.}.

Any truth about the universe can be derived from {\em Axiom Zero}. Like any false and undecidable propositions for that matter. So an additional principle is needed to derive the laws of the universe from {\em Axiom Zero}, and that is {\em the principle of logical consistency}. We select, among the possible logical consequences of {\em Axiom Zero}, only a logically consistent subset. That is, if the selected subset contains a proposition $q$, or if $q$ can be deduced from the other propositions it contains, it should not contain also its negation. This describes a possible universe. Any possible universe, including ours, can be obtained from {\em Axiom Zero} and the {\em principle of logical consistency}. So we may say that {\em Axiom Zero} is the ``mother of all possible worlds'', from which, effortlessly, any possible world appears, due to the {\em principle of logical consistency}.

But the {\em principle of logical consistency} does not tell what the laws are. We learn about the laws only by our observations, and, as Wheeler said, our observations can decide what the laws are. The outcome of each new observation is constrained to be consistent with the previous ones, so that the {\em principle of logical consistency} is not violated.

We arrive again at the conclusion that, to have {\em bits} which don't contradict one another, an underlying {\em it} which satisfies to those bits should exist.

\imagenocap{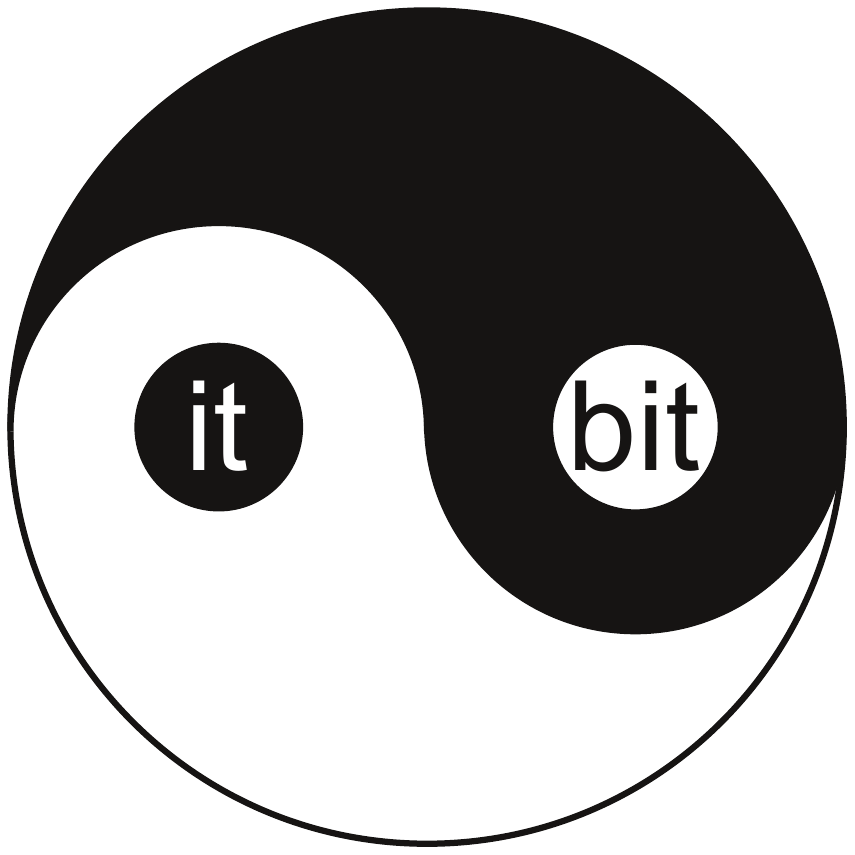}{0.37}


\end{document}